\documentclass[10pt,aps,prb,twocolumn,floatfix,floats,showpacs,superscriptaddress,raggedbottom]{revtex4-1}
\usepackage{epsfig}
\usepackage{graphicx,latexsym}
\usepackage{dcolumn}
\usepackage{amssymb,amsmath,bm}

\usepackage[dvips]{color}

\usepackage{hyperref}
\hypersetup{
    pdfnewwindow=true,       
    colorlinks=true,         
    linkcolor=blue,          
    citecolor=blue,          
    filecolor=magenta,       
    urlcolor=black           
}

\def\vec#1{{\bf#1}}
\def\eq#1{Eq.\ (\ref{#1})}
\def\mb#1{\mbox{\boldmath$#1$}}
\def\fig#1{Fig.\ \ref{#1}}

\begin{document}
\title{Double-finger-gate controlled spin-resolved resonant quantum transport \\
in the presence of a Rashba-Zeeman gap}
 \author{Chi-Shung Tang}
 \email{cstang@nuu.edu.tw}
 \affiliation{Department of Mechanical Engineering, National United University,
 Miaoli 36003, Taiwan}
 \author{Shu-Ting Tseng}
 \affiliation{Department of Electrophysics, National Chiao Tung University,
 Hsinchu 30010, Taiwan}
 \author{Vidar Gudmundsson}
 \affiliation{Science Institute, University of Iceland,
        Dunhaga 3, IS-107 Reykjavik, Iceland}
 \author{Shun-Jen Cheng}
 \email{sjcheng@mail.nctu.edu.tw}
 \affiliation{Department of Electrophysics, National Chiao Tung University,
 Hsinchu 30010, Taiwan}
 \affiliation{Physics Division, National Center for Theoretical Sciences, Hsinchu 300, Taiwan}



\begin{abstract}
We investigate double finger gate (DFG) controlled spin-resolved
resonant transport properties in an n-type quantum channel with a
Rashba-Zeeman (RZ) subband energy gap. By appropriately tuning the
DFG in the strong Rashba coupling regime, resonant state structures
in conductance can be found that is sensitive to the length of the
DFG system. Furthermore, a hole-like bound state feature below the
RZ gap and an electron-like quasi-bound state feature at the
threshold of the upper spin branch can be found that is insensitive
to the length of the DFG system.
\end{abstract}



\maketitle

\section{Introduction}

Spintronics utilizing the spin degree of freedom of conduction
electrons is an emerging field due to its applications from logic to
storage devices with high speed and very low power
dissipation.\cite{Loss1998,Zutic2004,Wolf2001} Manipulating the spin
information offers the possibility to scale down certain
semiconductor spintronic devices to the nanoscale and is favorable
for applications in quantum
computing.\cite{Awschalom2002,Awschalom2007,Heedt2012} Various
spin-orbit (SO) effects present in semiconductor structures provide
a promising way to spin manipulation in two-dimensional electron
gases (2DEG).\cite{Winkler2003,Meier2007}  Particularly, the Rashba
SO interaction is of importance in spintronic devices, such as the
gate-controllable spin field-effect
transistor.\cite{Datta1990,Bandyopadhyay2004,Koo2009,Tang2012,Sadreev2013}

The SO interaction can be induced when the transported electron
experiences a strong electric field due to an asymmetry in the
confinement potential, namely the Rashba SO interaction is caused by
a structure inversion asymmetry (SIA).\cite{Rashba60}  Especially,
the Rashba SO interaction due to SIA can be significantly induced in
2DEG confined by an asymmetric potential in semiconductor materials.
Experimentally, the Rashba interaction has been shown to be
effective for electron spin manipulation by using bias-controlled
gate contacts.\cite{Nitta1997} Recently, several approaches were
proposed to engineer a spin-resolved subband structure utilizing
magnetic
fields\cite{Muccio02,Brataas02,Zhang03,Wang03,Serra2005,Scheid2007}
or ferromagnetic materials.\cite{Sun03,Zeng03}  The combination of a
Rashba SO interaction and an external in-plane magnetic field may
modify the subband structure producing a spin-split Rashba-Zeeman
(RZ) subband gap feature.\cite{Pershin2004,Quay2010} To implement a
quantum information storing and transfer, not only coherent
manipulation\cite{Tang2012} but also resonant features involving SO
interactions are of importance.\cite{Zhang2014} This can be achieved
utilizing a double finger gate (DFG) forming a quantum dot in
between the fingers where electrons are subjected to the Rashba SO
coupling and the Zeeman interaction.

\begin{figure}[tbhq]
      \includegraphics[width=0.45\textwidth,angle=0]{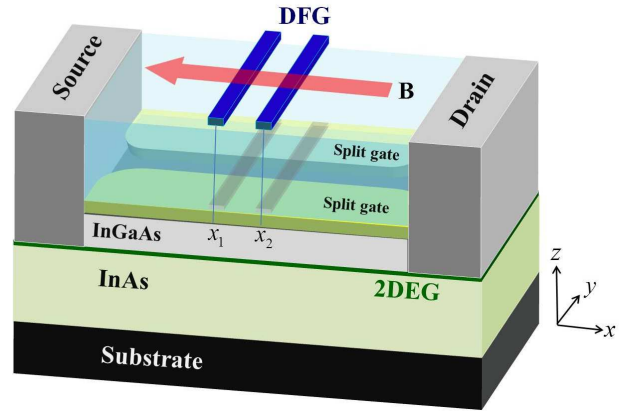}
      \caption{
(Color online) Schematic illustration of the quantum channel device
constructed with a 2DEG induced from InAs-In$_{1-x}$Ga$_x$As
semiconductor heterostructure. A split-gate is used to control the
channel width. An external in-plane magnetic field $\vec{B} = B
\hat{\mb{x}}$ ($B<0$). The DFG is consisted of two finger gates
located $x_1$ and $x_2$ to influence the spin-resolved resonant
quantum transport.} \label{fig1}
\end{figure}

In this work, we consider a split-gate induced narrow constriction
that is fabricated in a 2DEG in a narrow band gap semiconductor
heterostructure. A very asymmetric structure in the 2DEG leads to
strong SO coupling with the result that the Rashba effect is
dominant. We shall explore spin-resolved quantum transport
properties that are manipulated by a double finger gate (DFG) under
an external in-plane magnetic field as shown in \fig{fig1}.  Various
resonant transport mechanisms in the conductance will be
demonstrated analytically and numerically, including resonant states
(RS), hole-like bound states (HBS), and electron-like quasi-bound
states (EQBS).

The organization of the rest of this paper is as follows. In Sec.\
II we describe the propagation-matrix approach of tunneling through
a DFG system under in-plane magnetic field. In Sec.\ III we present
our calculated results on the spin-split subband structure and
the spin-resolved conductance. A concluding remarks is given in Sec.\
IV.


\section{DFG-contolled Transport Model}

In this section, we shall show how the split-gate confined quantum device
influenced by the RZ effect can be describe by a Hamiltonian
technique in order to obtain the spin-split subband structures.  The
corresponding group velocity and effective mass will be obtained to
analyze the spin-resolved resonant quantum transport behavior.  A
propagation matrix approach will be introduced to deal with the
DFG-controlled spin-resolved quantum transport.

\subsection{Hamiltonian of the DFG system}

As is illustrated for the device in \fig{fig1}, a two dimensional
electron gas (2DEG) is induced in an InAs-In$_{1-x}$Ga$_x$As
semiconductor heterojunction grown in the $(001)$ crystallographic
direction and is subjected to a split-gate voltage. A pair of
split-gates restrict the movement of the electrons of the 2DEG, and
therefore a quantum channel is generated in the $[100]$ direction.
Propagating electrons in the channel are driven from source to
drain.

In the absence of the finger gates, the transported electron is
affected by the Rashba effect $H_R$ due to SIA and the Zeeman effect
$H_Z$ induced by an external in-plane magnetic field, described by
the unperturbed Hamiltonian
\begin{equation}
\widetilde{H}_0 = H_0 + H_{\rm R} + H_{\rm Z}.
 \label{H0til}
\end{equation}
The first term describes a bare quantum channel that is described by
the ideal Hamiltonian
\begin{equation}
 H_0 = \frac{{\hbar ^2}{k^2}}{{2{m^*}}} + {U_c}(y).
\end{equation}
The first term is the kinetic energy of an electron in the 2DEG,
where $\hbar = h/2\pi$ is the reduced Planck constant.
A conduction electron has an assigned wave number $k$ satisfying $k^2=k_x^2 +
k_y^2$ and $m^*$ is its effective mass.  The second term is a
confining potential energy modeled by a hard-wall confinement
\begin{equation}
{U_c}(y) = \left\{ \begin{array}{l}
0, \, \left| y \right| < W/2\\
\infty, \, {\rm{otherwise}},
\end{array} \right.
\label{Uc}
\end{equation}
with $W$ being the width of the quantum channel that can be
controlled by applying a split-gate with negative voltage.

In the second term of \eq{H0til}, we consider a $(001)$
crystallographic 2DEG system, and hence the Rashba SO Hamiltonian
$H_{\rm R} = \alpha \left( \mb{\sigma}\times \mathbf{k}\right) \cdot
\hat{\mathbf{z}}$ couples the Pauli spin matrix $\mb{\sigma}$ to the momentum
$\mathbf{p}= \hbar \mathbf{k}$ can be reduced as a $k$-linear form
\begin{equation}
H_{\rm R} =  \alpha \left( \sigma_x k_y - \sigma_y k_x \right),
\label{HR}
\end{equation}
where the Rashba coupling strength $\alpha$ is proportional to the
electric field along $\hat{\mathbf{z}}$ direction perpendicular to the
2DEG.\cite{Nitta1997} The third term in \eq{H0til} describes an
applied external in-plane magnetic field  that is selected to be
antiparallel to the channel in the $[100]$ direction and has the form $\vec{B}
= B \hat{\mathbf{x}}$ ($B<0$). The longitudinal in-plane magnetic field
induced Zeeman term can be expressed as
\begin{equation}
H_Z = g{\mu_B}B \sigma_x,
\end{equation}
in which $g = g_s/2$ indicates half of the effective gyromagnetic
factor ($g_s = -15$ for InAs) and $\mu_B = 5.788\times
10^{-2}$~${\rm meV/T}$ is the Bohr magneton. In comparison with the
Zeeman Hamiltonian $H_Z$, we may rewrite \eq{HR} in a narrow channel
in the form $H_{\rm R} = g \mu_B B_R \sigma_y$, where the effective
Rashba magnetic field $\mathbf{B}_{\rm R}$ = $B_{\rm R}
\hat{\mathbf{y}}$ = $-\alpha k_x / (g\mu_B) \hat{\mathbf{y}}$.
Hence, the spin-resolved quantum channel system without the DFG may
be described by the unperturbed Hamiltonian
\begin{equation}
\widetilde{H}_0 = H_0 + g \mu_B \left( {B}{\sigma _x} + B_{\rm R}
\sigma_y \right) \, .
 \label{Htilde}
\end{equation}

In order to manipulate the spin-resolved resonant transport
properties, we applied the DFG on top of split gate with an
insulator in between, as shown in \fig{fig1}. We consider that the
width of the finger-gate scattering potential, $W$, should be less
than the Fermi wave length $\lambda_F = 31.4$~nm to be described as
a delta potential. We consider a high-mobility semiconductor
materials so that impurity effects can be neglected.  The considered
DFG system is then described by the scattering potential energy
\begin{equation}
 U_{\rm sc}(x) = e \sum_{j=1}^2 V_j \delta(x-x_j),
 \label{Usc}
\end{equation}
where $V_j$ indicates the bias potential applied by the finger gate
$j$.  The DFG system under investigation is thus described by the
Schr\"{o}dinger equation
\begin{equation}
\left[ \widetilde{H}_0 + U_{\rm sc}(x) \right] \Psi (x,y) = E \Psi (x,y).
\label{Htotal}
\end{equation}
The eigenfunction $\Psi (x,y)$ in \eq{Htotal} can be obtained by
summing over all occupied subbands, $n$, for the product of the
spatial wave functions and the spin states, given by
\begin{equation}
 \Psi (x,y) = \sum_n {\phi _n}(y){e^{i{k_x}x}}\chi_n \, .
 \label{Psi}
\end{equation}
Here the transverse wave function in subband $n$, of the split-gate
induced confining potential energy (\ref{Uc}), is of the form
 ${\phi _n}\left( y \right) = (\pi/W)^{1/2} \sin
 ( n\pi y / W )$
 with quantized bare subband energy
\begin{equation}
\varepsilon_{y,n} = \frac{\hbar^2}{2m^*}
 \left( \frac{n\pi}{W}
 \right)^2 .
 \label{ey}
\end{equation}
After some algebra, the corresponding eigenenergies of
(\ref{Htotal}) can be obtained
\begin{equation}
E_n^\sigma(k_x) = \frac{\hbar^2 k_x^2}{2 m^*} + \varepsilon_{y,n} +
\sigma g \mu_B B_{\rm RZ},
\label{En}
\end{equation}
where $\sigma = \pm$ is the spin index, and $B_{\rm RZ}^2 = B^2
+B_{\rm R}^2$ is the effective RZ magnetic field with $B_{\rm R} = 2
\alpha k_x /(g\mu_B)$ being a momentum dependent magnetic field due
to the Rashba effect. This expression indicates that the subband
spin-split energy gap $\Delta E_{\rm RZ}$ = $E_n^+ - E_n^-$ =
$2g\mu_B B_{\rm RZ}$ can be changed by tuning the effective RZ
magnetic field. It is interesting to note that this spin-split
energy gap $\Delta E_{\rm RZ}$ is reduced to $\Delta E_{\rm Z} =
2g\mu_B B$ in the zero momentum limit.

For simplicity, we employ the Fermi-level in a 2DEG as an energy
unit, namely $E^*$ = $E_F$ = $\hbar^2 k_F^2/2m^*$ with $m^\ast$ and
$\hbar$ being, respectively, the effective mass of an electron and
the reduced Planck constant. In addition, one selects the inverse
wave number as a length unit, namely  $l^* = k_F^{-1}$.
Correspondingly, the magnetic field is in units of $B^* = \mu_B^{-1}
E^*$, and the Rashba SO-coupling constant $\alpha$ is in units of
${\alpha}^* = E^*l^*$.  In the following we consider a sufficient
narrow channel by assuming the channel width $W=\pi l^*=15.7$~nm so
that the bare subband energy due to $U_c(y)$ is simply
$\varepsilon_{y,n} = n^2$. The energy dispersion can thus be
expressed as
\begin{equation}
E_n^{\sigma}  = k_x^2 + n^2 + \sigma \sqrt{ (gB)^2 + (2 \alpha
k_x)^2 },
 \label{En_dimless}
\end{equation}
where $\sigma = \pm$ indicates the upper ($+$) and lower ($-$) spin
branches.  Sufficiently low temperature $k_B T < 0.1 \Delta
\varepsilon$ or $T < 23$~K is required to avoid thermal broadening
effect.


\subsection{Spin-resolved quantum transport}

In order to investigate the DFG-controlled spin-resolved quantum
transport properties, we shall explore how the spin-mixing effect
due to the RZ coupling influences the propagating and evanescent
modes for a given energy of an incident electron.  The energy dispersion
relation (\ref{En_dimless}) can be rewritten in the form
\begin{equation}
 k_x^4 - \left[ 4\alpha^2 - \left( {K_n^\sigma} \right)^2 \right]{k_x^2}
 +  \left( {K_n^\sigma} \right)^2 - (gB)^2 = 0,
\end{equation}
where $K_n^\sigma = E_n^\sigma - n^2$ indicates the ideal kinetic
energy of an electron in the transverse subband $n$ in the absence
of a RZ effect.  To proceed, one has to label the four longitudinal
wave numbers $k_x$ as the right-going $k_{\sigma}$ and left-going
$q_{\sigma}$, in which the notation $\sigma = +$ indicates spin-up
mode and $\sigma = -$ stands for spin-down mode.

Below, we focus on a sufficiently narrow quantum channel to explore
the first two conductance steps associated with the two spin
branches of a transported electron occupying the lowest subband.  We
calculate the quantum transport properties by using a generalized
spin-resolved propagation matrix method, in which the spin branches
as well as spin-flip scattering mechanisms are taken into account.
The energy dispersion shown in \fig{fig3}(a) essentially divides the
energy spectrum into three regimes, namely the low energy regime
$E^-_{\rm bottom} < E < E^-_{\rm top}$, the intermediate energy
regime $E^-_{\rm top} < E < E^+_{\rm bottom}$, and the high energy
regime $E > E^+_{\rm bottom}$.   In the low and high energy regimes,
there are four propagating modes with real $k_{\sigma}$ and real
$q_{\sigma}$.  It should be noted that there are two propagating and
two evanescent modes in the intermediate energy regime or the RZ
energy gap region where the evanescent modes manifest a bubble
behavior with imaginary wave vectors.\cite{Tang2012}

The spin-resolved wave functions around the scattering potential
$U_{\rm sc}$ located at $x_j$ given by \eq{Usc} can be formally
expressed as
\begin{equation}
\psi \left( x \right) = \sum\limits_{\sigma = \pm } {{A_\sigma
}{e^{i{k_\sigma }x}}\chi ({k_\sigma })}  + \sum\limits_{\sigma  =
\pm } {{B_\sigma }{e^{i{q_\sigma }x}}\chi ({q_\sigma })} ,\quad x <
x_j \label{wf1}
\end{equation}
\begin{equation}
\psi \left( x \right) = \sum\limits_{\sigma = \pm } {{C_\sigma
}{e^{i{k_\sigma }x}}\chi ({k_\sigma })}  + \sum\limits_{\sigma  =
\pm } {{D_\sigma }{e^{i{q_\sigma }x}}\chi ({q_\sigma })} ,\quad x >
x_j \label{wf2}
\end{equation}
where $A_\sigma$ and $C_\sigma$ indicate the right-going wave
amplitude corresponding to $k_\sigma$, while $B_\sigma$ and
$D_\sigma$ represents the left-going wave amplitude corresponding to
$q_\sigma$, and $\chi_\sigma$ stands for the momentum dependent spin
states.  It is possible to obtain the propagation matrix equation by
matching suitable boundary conditions as shown below around
the free space or the scattering potential induced by the finger
gates, namely the electronic wave function is continuous
\begin{equation}
\psi \left( {x_j^- } \right) = \psi \left( {x_j^+} \right)
 \label{bc1}
\end{equation}
and the derivative of wave function is discontinuous by a deduction
of delta scattering potential energy, given by
\begin{equation}
\psi '\left( {x_j^-} \right) = \psi '\left( {x_j^+} \right) -
e{V_j}\psi \left( {x_j^+} \right)\, .
 \label{bc2}
\end{equation}

Before matching the above boundary conditions, it is convenient to
define the reflection coefficient $r_{{\sigma_i},{\sigma_f}} =
B_{\sigma_f} / A_{\sigma_i}$ and the transmission coefficient
$t_{{\sigma_i},{\sigma_f}} = C_{\sigma_f} / A_{\sigma_i} $ that
involves the spin flip states ($\sigma_i \neq \sigma_f$) and spin
non-flip states ($\sigma_i = \sigma_f$).  Taking into account the
possible incident spin states $\sigma$ and $\bar{\sigma }$ allows us
to write the propagation matrix equation (PME) in terms of the total
propagation matrix $\rm{\bf P}^{\rm T}$
\begin{equation}
\left[ {\begin{array}{*{20}{c}}
   1 & 0  \\
   0 & 1  \\
   {r_{\sigma ,\sigma }} & {r_{\bar \sigma ,\sigma }}  \\
   {r_{\sigma ,\bar \sigma }} & {r_{\bar \sigma ,\bar \sigma }}  \\
\end{array}} \right] = {{\bf{P}}^{\rm T}}\left[ {\begin{array}{*{20}{c}}
   {t_{\sigma ,\sigma }} & {t_{\bar \sigma ,\sigma }}  \\
   {t_{\sigma ,\bar \sigma }} & {t_{\bar \sigma ,\bar \sigma }}  \\
   0 & 0  \\
   0 & 0  \\
\end{array}} \right]\, .
 \label{PME}
\end{equation}
To proceed, we match the wave functions \eq{wf1} and \eq{wf2} using
the boundary conditions \eq{bc1} and \eq{bc2} corresponding to
$U_{\rm sc}(j)$, and then we rearrange these equations into a $4
\times 4$ interface propagation matrix ${\bf{P}^\delta}(j)$ of the
delta scattering potential $j$.  Moreover, one has to construct a
spin-resolved free-space propagation matrix ${\bf{P}^{\rm F}}(L)$
with length $L$ between the finger gates, given by ${\bf{P}^{\rm
F}}(L) = \exp (-ik_{i,j}L) \delta_{i,j}$, in which $k_{1,1}=
k_{\sigma}$, $k_{2,2}= k_{\bar \sigma}$, $k_{3,3}= -q_{\sigma}$, and
$k_{4,4}= -q_{\bar \sigma}$.  The total propagation matrix
$\mathbf{P}^{\rm T}$ thus consists of the matrices for the first and
second scattering delta potentials $\mathbf{P^\delta }(1)$ and
$\mathbf{P}^\delta(2)$ induced by the DFG as well as a free space
propagation matrix $\mathbf{P}^{\rm F}(L)$ between them, given by
\begin{equation}
\mathbf{P}^{\rm T} = \mathbf{P^\delta }(1) \mathbf{P}^{\rm F}(L)
 \mathbf{P}^\delta(2)\, .
\end{equation}
Solving the PME numerically, we may obtain the reflection and
transmission coefficients of the scattering intermediate and final
states in the presence of the DFG.

We consider an electron injected from the left reservoir (source
electrode) and transported to the right reservoir (drain electrode)
for a given incident energy.  Solving for the spin non-flip and flip
reflection coefficients $r_{\sigma ,\sigma}$ and $r_{\sigma ,\bar
\sigma}$, as well as the spin non-flip and flip transmission
coefficients $t_{\sigma ,\sigma}$ and $t_{\sigma ,\bar \sigma}$, we
can calculate numerically the conductance based on the
Landauer-B\"{u}ttiker framework\cite{Landauer1970,Buttiker1990}
\begin{equation}
G = G_0 \sum\limits_{\sigma_L,\sigma_R}
 \frac{v_{\sigma_R}}{v_{\sigma_L}}
 \left| t_{\sigma_L,\sigma_R} \right|^2 \, .
\label{eq3.2.30}
\end{equation}
Here $G_0$ = $e^2/h$ is the conductance quantum per spin branch, and
$\sigma_L$ and $\sigma_R$ indicate, respectively, the spin branches
of the incident and transmitted waves in the left and right leads.
Therefore, ${v_{{\sigma_L}}}$ and ${v_{{\sigma_R}}}$ represent the
group velocity of corresponding modes in the left and right
reservoirs, respectively.


\section{Numerical Results}

Calculations presented below are carried out under the assumption
that the electron effective mass $m^{\ast}=0.023 m_0$, which is
appropriate for the InAs-In$_{1-x}$Ga$_x$As semiconductor interface
with the typical electron density $n_e \sim
10^{12}$~cm$^{-2}$.\cite{Nitta1997} Accordingly, the energy unit is
$E^*$ = 66~meV, the length unit $l^*$ = 5.0~nm, the magnetic field
unit $B^* = 1.14$~$\textrm{kT}$, and the spin-orbit coupling
parameter is in units of ${\alpha^*} = 330$~meV$\cdot$nm.  In
addition, the bias potential of the finger gate is in units of $V^*
= 330$~mV$\cdot$nm. By using the above units, all physical
quantities presented below are dimensionless.\cite{Tang2012}


\subsection{Subband structures with Rashba-Zeeman effect}

\begin{figure}
\centerline{\includegraphics[width = 0.42 \textwidth]{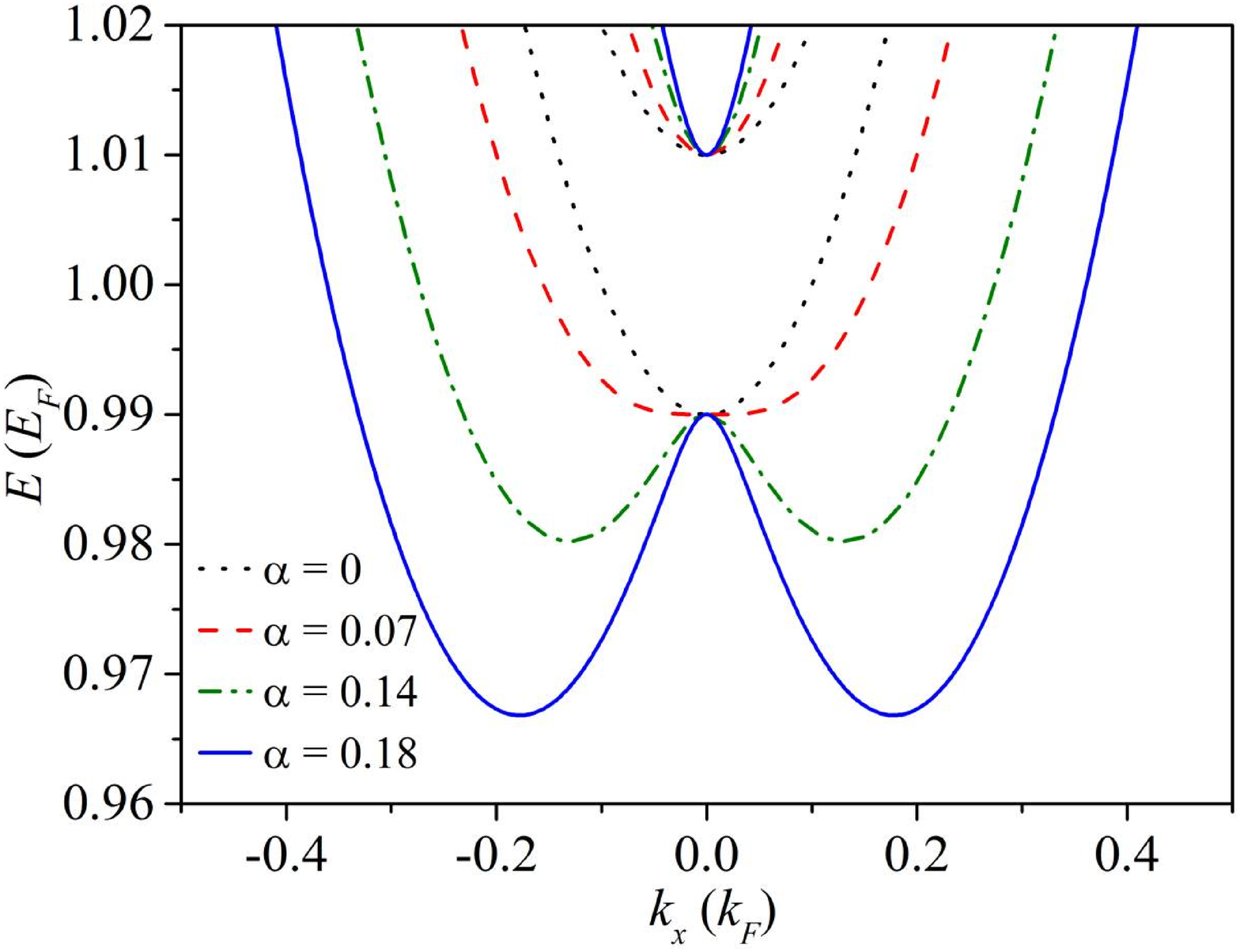}}
\caption{(Color online) Subband structure in the presence of
Rashba-Zeemand interaction under an in-plane magnetic field $gB =
0.01$ for the Rashba coefficients $\alpha$ = 0 (black dot), 0.07
(red dash), 0.14 (green dash dot), and 0.18 (blue solid).}
 \label{fig2}
\end{figure}

It is known that the presence of an in-plane magnetic field may
split the spin degenerate parabolic energy dispersion vertically
toward the higher and lower energy and manifests an energy
difference $\Delta {E_{\rm{Z}}} = 2gB$, as shown by black dotted
line in \fig{fig2}. In addition, the Rashba SO coupling may let the
subband structure shift horizontally toward the positive and
negative momentum directions.

By appropriately tuning the applied in-plane magnetic field, the
Rashba SO interaction can be separated into several coupling
regimes.

In the intermediate Rashba coupling regime $2\alpha^2 = gB$, the
spin-up branch is still parabolic while the spin-down branch
manifests a flat subband bottom and $\Delta E_{\rm RZ} = \Delta
E_{\rm Z}$, as shown by red dashed line in \fig{fig2}.  In the
strong Rashba coupling regime $gB < 2\alpha^2 \leq 4gB$, the
combination of the Rashba and Zeeman interactions provide a
possibility to generate a RZ gap with a significant subband in the
spin-down branch, as shown by green dash-dotted line in \fig{fig2}.
A significant zero point energy of a transported electron in the DFG
system occurs at the subband top of the spin-down branch.
Furthermore, we shall show below that, in the ultra-strong coupling
regime $\alpha^2 > 2gB$, the zero point energy will be changed to
the subband bottom of the spin-down branch.

\begin{figure}
\centerline{\includegraphics[width = 0.4 \textwidth]{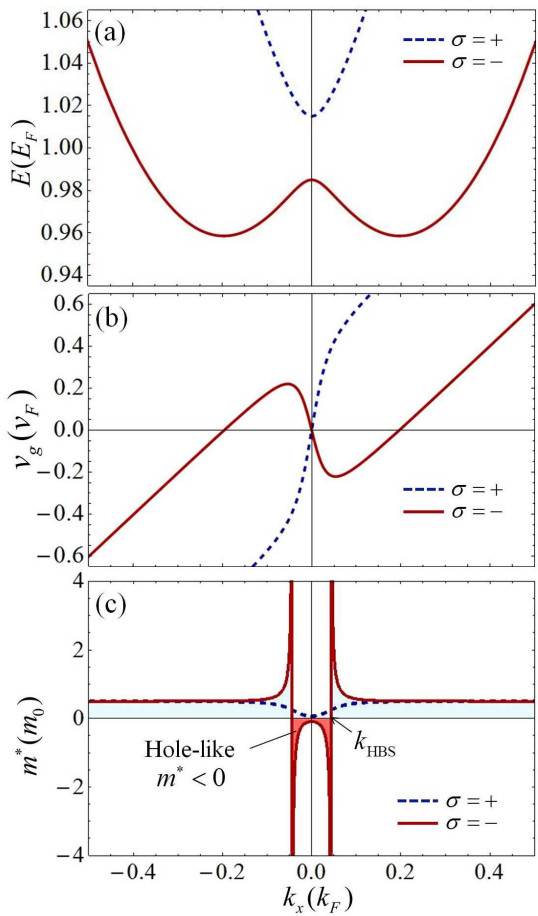}}
\caption{(Color online) (a) The first spin-split subband structure
with Rashba coefficient $\alpha$ = 0.2 and magnetic field $gB$ =
0.015. The spin-up branch (blue dash) is parabolic while the
spin-down branch (red solid) manifests a top and two bottoms of the
same energy. (b) Corresponding group velocity of spin-up (blue dash)
and spin-down (red solid) branches. (c) Corresponding effective mass
in momentum space.  The effective mass of a spin-up electron is
always positive (blue dash). However, the effective mass of
spin-down electron is positive for large wave number $k$ (blue
shadow) but is negative for small $k$ (red shadow) due to the strong
RZ effect.}
 \label{fig3}
\end{figure}

In \fig{fig3}(a), we show the energy dispersion of the first subband
with the Rashba coefficient $\alpha = 0.2$ and an in-plane magnetic field
$gB$ = 0.015. This is within the strong Rashba coupling regime,
$2\alpha^2/(gB) > 1$. The subband bottom of the upper spin branch is
at $E_{\rm bottom}^+ = 1 + gB$. However, the subband bottom at $k_x
= 0$ of the lower spin branch becomes a subband top with the same
energy $E_{{\rm top}}^- = 1 - gB$. Therefore, the RZ energy gap of
the plus and minus branches $\Delta {E_{\rm{RZ}}}$ is exactly the
Zeeman energy $\Delta {E_{\rm{Z}}}$.

In order to explore the spin-resolved transport properties, it is
important to define the group velocity of an electron in the
$\sigma$ spin branch
\begin{equation}
v_{\sigma} = \frac{dE_n^{\sigma}}{d{k_x}} = 2{k_x} + \sigma
\frac{{4{\alpha^2}{k_{x}}}}{\sqrt {(gB)^2 + 4{\alpha ^2}k_x^2}}
\label{vg}
\end{equation}
as shown in \fig{fig3}(b).  Defining the velocity allows us to
determine a local minimum and a maximum in the subband
structures by setting the group velocity identically zero.  We
see that there are two subband bottoms in the lower spin branch at
$k_x = \pm \left[ \alpha^2 - (gB/2\alpha)^2 \right]^{1/2}$ with the
same energy $E_{\rm bottom}^- = 1 - \left[ \alpha^2
+(gB/2\alpha)^2\right]$.

To identify an electron-like ($m^* > 0$) and a hole-like ($m^* <
0$) nature, it is necessary to define the effective mass by
performing second derivation of energy band, given by
\begin{equation}
 \frac{1}{m^*_{\sigma}}
 = \frac{d^2 E_n^\sigma}{dk_x^2}
 = 2 + \sigma \frac{4\alpha^2\left(gB\right)^2}
 {\left[ {\left(gB\right)^2 + \left(2\alpha k_x\right)^2} \right]^{3/2}}\, .
 \label{effectivemass}
\end{equation}
This expression allows us to define hole-like bound states (HBS)
that occurs when the effective mass goes to infinity, as shown in
\fig{fig3}(c). The corresponding HBS wave number can be analytically
expressed as
\begin{equation}
 k_{\rm HBS} = \sqrt{
 \left|
 {\left[ {\frac{\left(gB\right)^2}{4\alpha}} \right]}^{2/3}
 - \left( \frac{gB}{2\alpha }\right)^2
 \right| }\, .
  \label{k_HBS}
\end{equation}
The fact that $k_{\rm HBS}$ goes to zero if $2\alpha^2 = gB$ implies
the HBS feature can be found only in the strong Rashba coupling
regime $2\alpha^2 > gB$.

It is clearly sown in \fig{fig3}(c) that the effective mass is
always positive in the spin-up branch (blue dashed line) while the
effective mass in the spin-down branch (red solid line) is allowed
to be negative in the small momentum regime  $|k_x| < k_{\rm HBS}$
(red shadow). The corresponding HBS energy can be obtained as
\begin{equation}
 E_{\rm HBS} = 1 - \left( \frac{gB}{2\alpha} \right)^2
  +\left[ \frac{\left(gB\right)^2}{4\alpha} \right]^{2/3}
  - \left[ 2 \alpha^2  \left(gB\right)^2 \right]^{1/3}
 \label{E_HBS}
\end{equation}
to investigate the HBS in the conductance as we shall demonstrate in the
next section.  Having a finite group velocity but an infinite effective
mass implies that the electron will be restricted to the energy
level corresponding to the inflection point in energy. This is
recognized as a HBS in the lower spin branch. The HBS nature will
significantly influence the spin-resolved resonant quantum transport
behavior.


\subsection{DFG controlled transport}

In this section, we discuss how the conductance is influenced by the
DFG to manifest various electron-like and hole-like peak structures
due to the presence of the RZ coupling.  The length $L$ between
the two finger gates is tuned to demonstrate these spin-resolved
quantum transport features.

\begin{figure}[t]
\includegraphics[width = 0.48 \textwidth, angle=0] {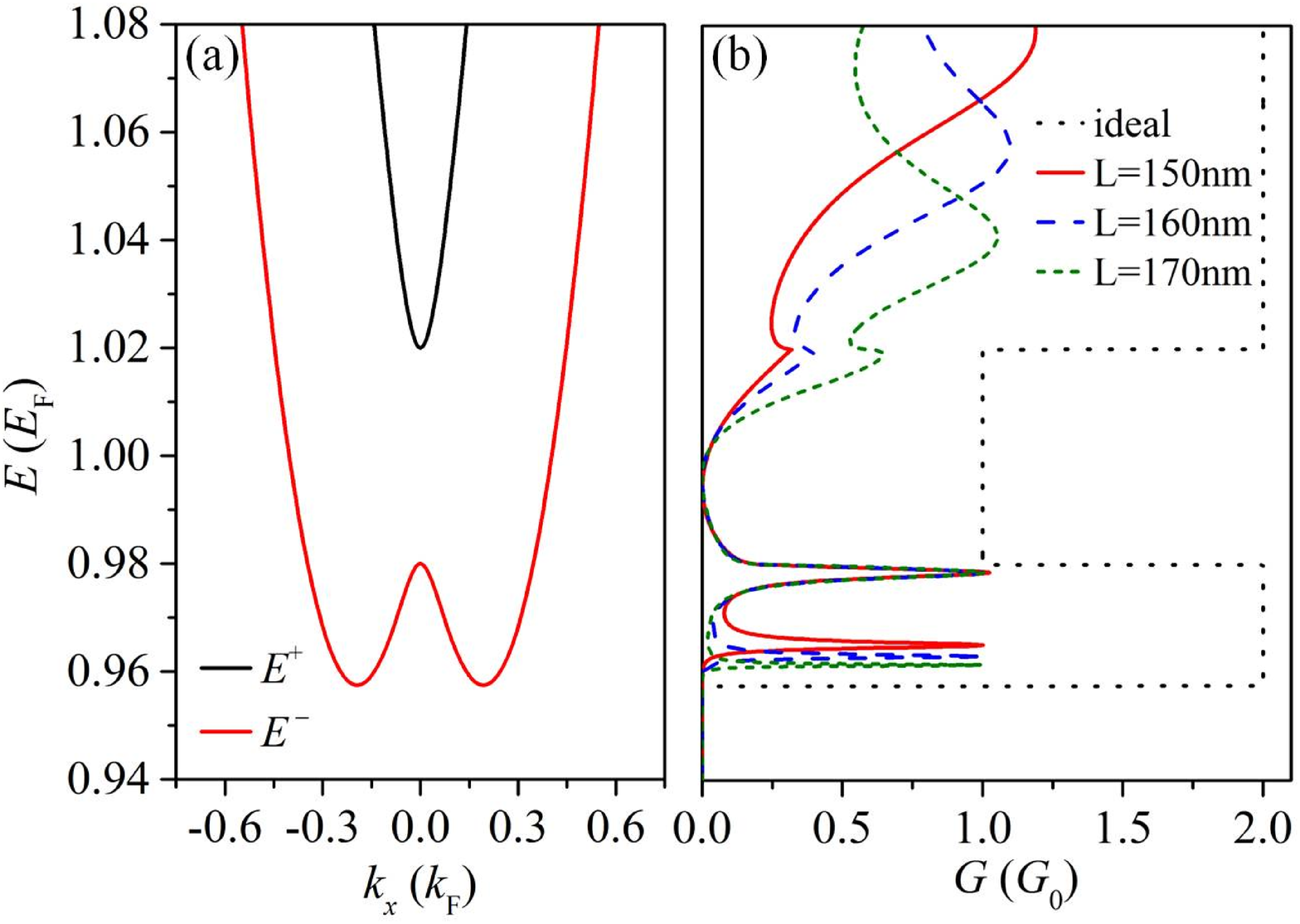}
\caption{(Color online) (a) Energy dispersion and (b) corresponding
conductance as a function of incident electron energy with different
length $L$ between the two finger gates. The ideal case without DFG
is shown by black dotted line. The distance between the gate fingers
$L$ is selected as $150$~nm (red solid), $160$~nm (blue dashed),
$170$~nm (green short dashed).  Other parameters are $\alpha$ = 0.2,
$gB$ = 0.02, and $V_1$=$V_2$=0.6.} \label{fig4}
\end{figure}

Figure \ref{fig4} shows the spin-split energy dispersion and its
corresponding influence on the conductance. Obvious
are the peaks corresponding to the resonant ground state in low energy
regime and the first excited state in the high energy regime. In
\fig{fig4}(a), we show the spin-split energy dispersion by taking
the Rashba coefficient $\alpha$ = 0.2 (66 meV nm) and $gB$ = 0.02
($B=3$~${\rm T}$) to ensure that the system is in the strong SO
coupling regime ($2\alpha^2 >gB$). The upper spin branch $E^+$
manifests a single band bottom $E^+_{\rm
bottom}$=$\varepsilon_{y,1}+gB$=$1.02$. The lower spin branch $E^-$
exhibits a single band top at energy $E^-_{\rm
top}$=$\varepsilon_{y,1}-gB$=$0.98$ and two band bottoms with the
same energy $E_{\rm bottom}^- = \varepsilon_{y,1} - \left[ \alpha^2
+(gB/2\alpha)^2\right]$=$0.958$.

In \fig{fig4}(b), we demonstrate how the transport properties are
affected by the applied DFG by fixing the finger gate voltage
$V_1$=$V_2$=0.6 while tuning the length $L$ between the two finger
gates. In the low kinetic energy regime $E_{\rm bottom}^- < E <
E^-_{\rm top}$, there are two different resonant features in
conductance. The first resonant feature at a lower energy is a
resonant state (RS) due to multiple scattering between the two
finger gates.  When the transported electron is in the double
scattering potential induced by the finger gates, it is
quasi located in an imaginary quantum well embedded in the quantum
channel. The $m$th RS states are sensitive to the length $L$ between
the finger gates and can be approximately estimated by the
theoretical formula
\begin{equation}
E_{{\rm RS},m}^{\rm th} = E_{\rm zero} + \varepsilon_{x,m}\, ,
 \label{ERS}
\end{equation}
in which ${\varepsilon _{x,m}} = ( m\pi/L)^2$ is the $m$th energy
level due to the DFG with zero point energy $E_{\rm zero}$.  When
the Rashba coupling strength is within the ultra-strong coupling
regime $\alpha^2 > 2gB$ as shown in \fig{fig4}, the zero point
energy $E_{\rm zero}$ is identically the subband bottom of the
spin-down branch $E_{\rm bottom}^-$.  Theoretically, the first RS
structures in conductance are at $E_{{\rm RS},1}^{\rm th} = E_{\rm
bottom}^- + \varepsilon_{x,1}$ = 0.968, 0.966, 0.965 for $L=150,
160, 170$~nm, respectively. In \fig{fig4}(b), the numerical
calculation by means of propagation matrix method gives $E_{{\rm
RS},1}$= 0.965, 0.963, and 0.961 for $L=150, 160, 170$~nm,
respectively. To estimate the accuracy of our theoretical
estimation, we define the mean absolute percentage error (MAPE) in
energy as
\begin{equation}
M = \frac{100\%}{n}\sum_{i=1}^n
 \left|\frac{E_{L_i} - E^{\rm th}_{L_i}}{E_{L_i}}\right|\, ,
  \label{MAPE}
\end{equation}
where $n$ is the number of selected lengths $L_i$ of the DFG system.
This formula gives the MAPE of the RS structure in conductance to be
$M_{{\rm RS,1}} = 0.36\%$. Similarly, the theoretical estimation of
the fourth RS structures in the conductance are $E_{{\rm RS},4}^{\rm th}
= E_{\rm bottom}^- + \varepsilon_{x,4}$ = 1.118, 1.098, 1.082 for
$L=150, 160, 170$~nm, respectively. In the high kinetic energy
regime $E > E_{\rm bottom}^+$, we can find RS peaks in the conductance
at $E_{{\rm RS},4}$ = 1.078, 1.057, and 1.04 for $L=150, 160,
170$~nm, respectively. The MAPE of the fourth RS peak in the conductance
is $M_{{\rm RS},4} = 3.86\%$.

\begin{figure}[h]
\includegraphics[width = 0.4 \textwidth,angle=0] {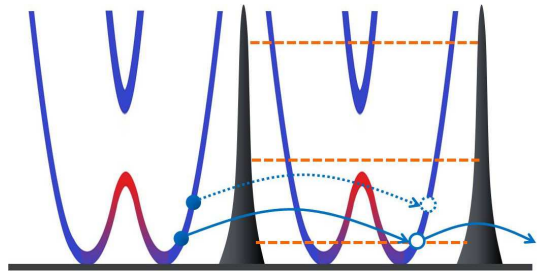}
\caption{(Color online) Schematic illustration of resonant state
(RS) enhanced transport (solid blue arrow) if the incident
electronic energies coincide with the resonant energy levels with
zero-point energy at the subband bottom of the spin-down branch, as
is shown by broken orange line. However, the electron transmission
is not allowed if the incident electron energy is not on the RS
shown by the dashed blue arrow.}
 \label{fig5}
\end{figure}

The transport mechanisms of these conductance peaks are schematically
shown by solid blue arrows in \fig{fig5}. These conductance peaks
are associated with resonant bound energy levels
$\varepsilon_{x,m}$ and can be tuned by changing the length $L$
between the two finger gates. They will be closer to the lower
subband bottom when the length $L$ is increased.  We note in passing
that the second and the third RS structures $E_{{\rm RS},2}$ and
$E_{{\rm RS},3}$ can not be found in the conductance, these RS features
are suppressed due to the formation of the RZ energy gap.

\begin{figure}[h]
\includegraphics[width=0.4\textwidth,angle=0] {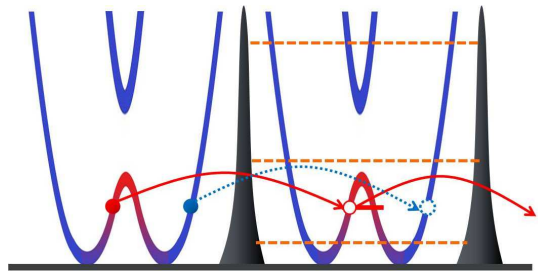}
\caption{(Color online) (HBS) Schematic illustration of a hole-like
particle transport. The electron transmission is not allowed if the
electron energy is not aligned with the resonant state (blue dashed
arrow). However, when the electron with incident energy equals the
bound state energy of the hole-like subband structure (red solid
line), it may contribute to a length insensitive peak structure in
conductance (red solid arrow).}
 \label{fig6}
\end{figure}

The second resonant feature in the conductance shown in \fig{fig4}
is a hole-like bound state (HBS) at the same energy $E_{\rm HBS} =
0.978$ for $L=150, 160, 170$~nm. The corresponding theoretical
prediction based on \eq{E_HBS} is given by $E_{\rm HBS}^{\rm th} =
0.972$. The corresponding MAPE is $M_{\rm HBS} = 0.62\%$.  It is
found that such HBS structure in the conducatance is independent of
the distance $L$ between the two finger gates. In the intermediate
kinetic energy regime (i.e.\ in the RZ gap energy regime), a small
peak in the conductance can be found at the threshold of the upper
spin branch. This structure is recognized as a electron-like
quasi-bound state (EQBS).  In comparison with the case of a single
finger gate system,\cite{Tang2012} the EQBS feature is a peak
structure instead of dip structure in conductance.

This HBS mechanism is schematically shown by red arrows in
\fig{fig6} indicating an electron occupying an inner mode  in the
low  kinetic energy regime $E_{\rm bottom}^- < E < E^-_{\rm top}$
forming a HBS below the subband top of the spin-down branch.
However, the electron with energy $E_{\rm HBS}$ occupying the outer
mode is at off-resonant energy and cannot be transmitted through the
DFG system, as is shown by the blue dashed arrows in \fig{fig6}.

\begin{figure}[htb]
\includegraphics[width=0.45\textwidth,angle=0] {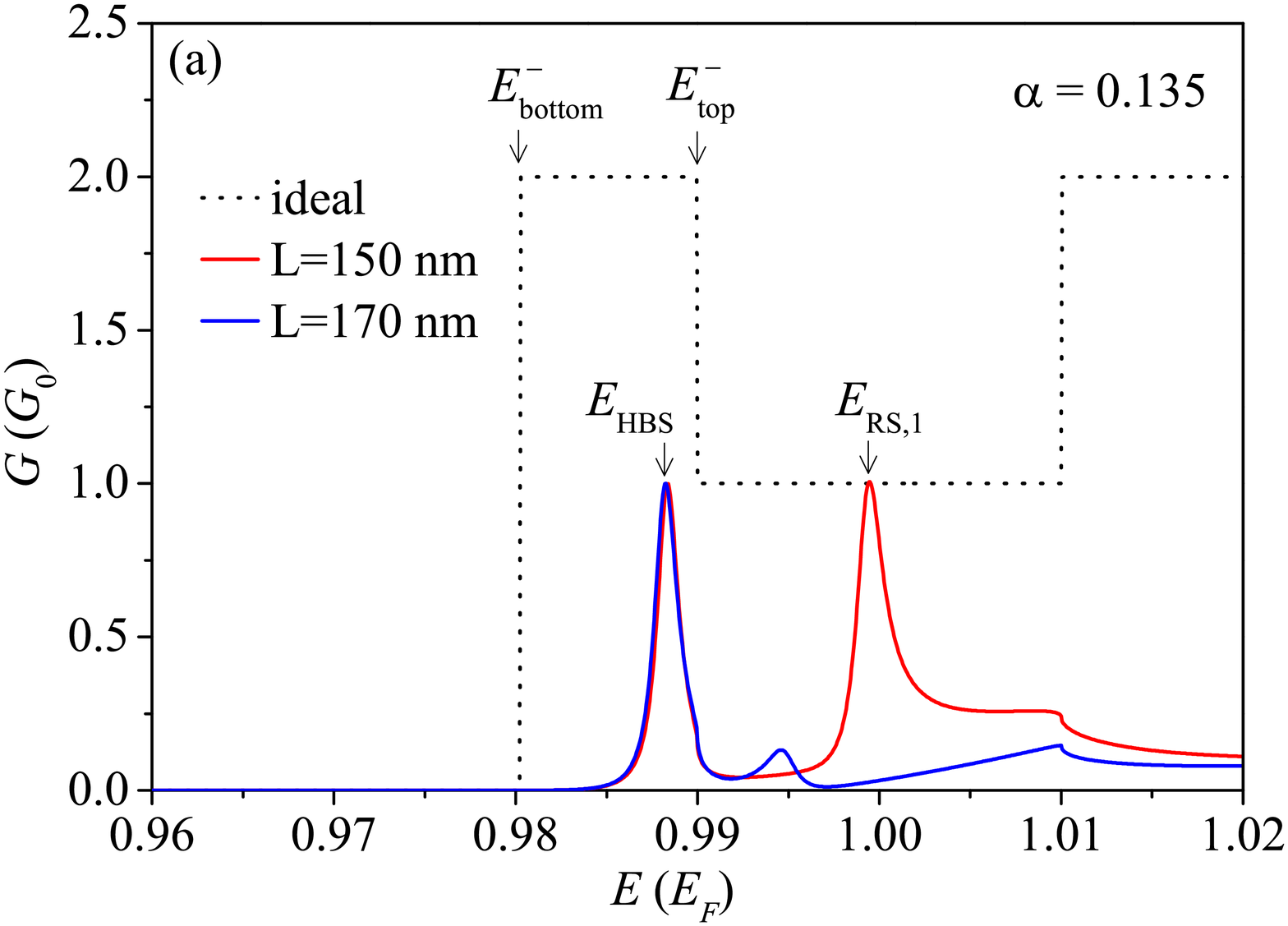}
\includegraphics[width=0.45\textwidth,angle=0] {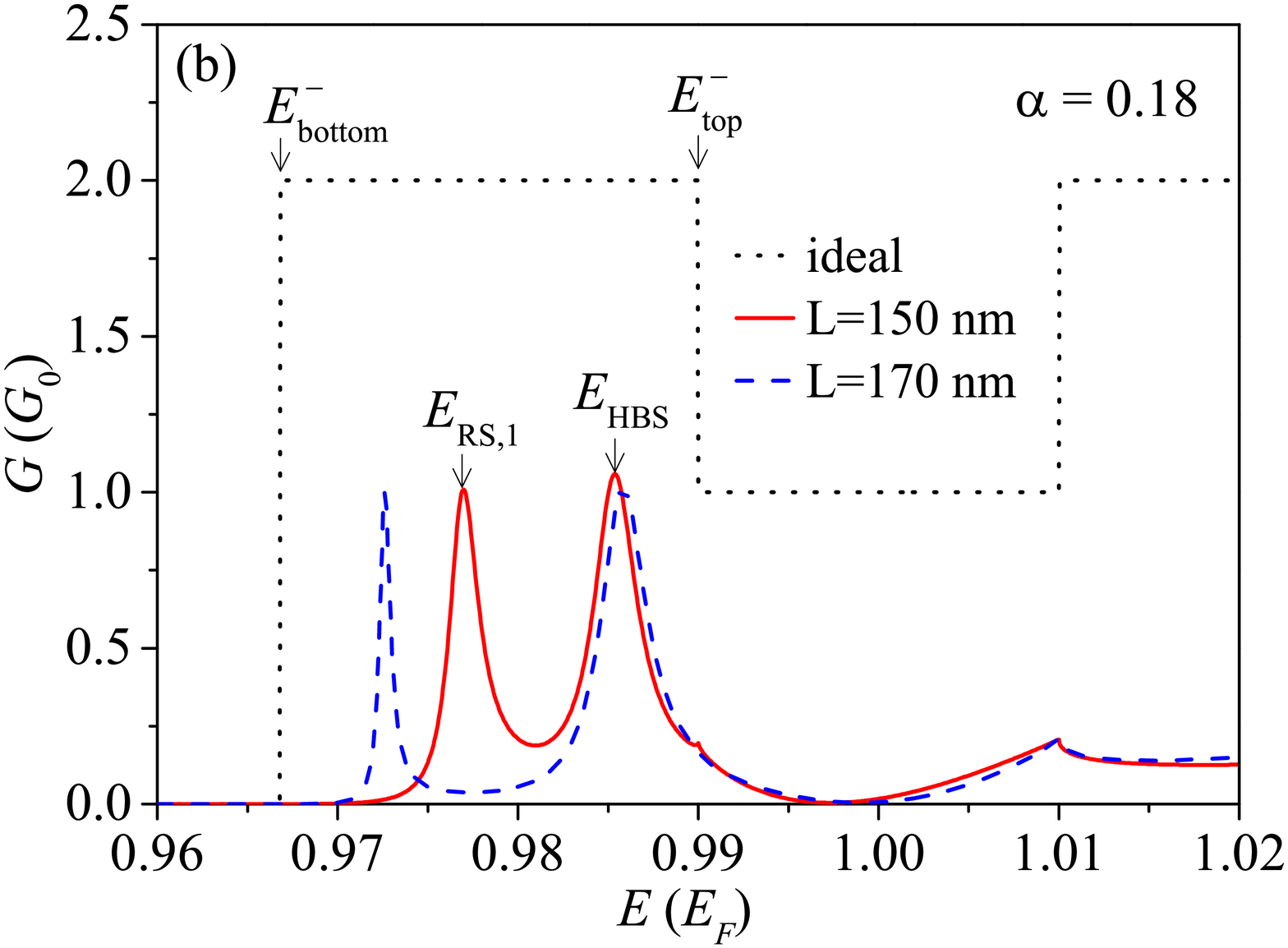}
\caption{(Color online) Conductance is plotted as a function of the
incident electron energy under magnetic field strength $gB = 0.01$:
(a) $\alpha$=0.135 and (b) $\alpha$=0.18 with length $L$ = 150 nm
(red solid) and 170 nm (blue dash) between the two finger gates. The
DFG system is subject to the same positive potential $V_1$ = $V_2$ =
0.6 in both cases.}
 \label{fig7}
\end{figure}

In \fig{fig7}, we show the conductance as a function of energy in an
in-plane magnetic field $gB$ = 0.01 ($B$ = 1.5~T) while tuning the
Rashba coefficient to be \fig{fig7}(a) $\alpha$ = 0.135 within the
strong coupling regime and \fig{fig7}(b) $\alpha$ = 0.18 within the
ultra-strong coupling regime. In both cases, we compare results for
the distance $L$ = 150 nm (red solid line) and $L$ = 170 nm (blue dashed line)
between the two finger gates.

In the strong Rashba regime as shown in \fig{fig7}(a), since the
energy difference between the subband top $E^-_{\rm top}$ and the
subband bottom $E^-_{\rm bottom}$ of the spin-down branch is small
the transported electron occupying the RS manifests a conductance
peak at $E_{\rm RS,1}$ satisfying \eq{ERS}.  Our theoretical
estimation predicts the zero point energy of the RS peaks in the
conductance is at the subband top of the spin down branch, namely
$E_{\rm zero} = E^{-}_{\rm top}$ = 0.9900.  Therefore, we can
estimate that the first RS peak in the conductance can be found at
energy $E_{\rm RS,1}^{\rm th}$ = 1.0000 and 0.9978 for $L$ = 150 and
170~nm, respectively. The numerical result shown in \fig{fig7}(a)
gives $E_{\rm RS,1}$ = 0.9997 and 0.9945 for $L$ = 150 and 170~nm,
respectively.  The MAPE of the first RS state $M_{{\rm RS},1} =
0.17\%$ in the case of $\alpha = 0.135$ is very accurate.

In the ultra-strong Rashba regime shown in \fig{fig7}(b), the energy
difference between the subband top $E^-_{\rm top}$ and the subband
bottom $E^-_{\rm bottom}$ of the spin-down branches become
substancial. Therefore, the zero point energy of the first RS peak
in the conductance satisfying \eq{ERS} will be changed to be $E_{\rm
zero}$ = $E^-_{\rm bottom}$ = 0.9668, and the theoretical estimation
of the first RS peak is $E_{\rm RS,1}^{\rm th}$ = 0.9768 and 0.9746
for $L$ = 150 and 170~nm respectively. The numerical result shown in
\fig{fig7}(a) gives $E_{\rm RS,1}$ = 0.9769 and 0.9726 for $L$ = 150
and 170~nm, respectively.  The MAPE of the first RS state $M_{{\rm
RS},1} = 0.11\%$ in the case of $\alpha = 0.18$ is very
accurate.

In summary, the above results shown in \fig{fig7} demonstrate that when
the Rashba coupling is increased from the strong to the ultra-strong
regime, the zero point energy of the first RS peak in the conductance
will be changed from $E^-_{\rm top}$ to $E^-_{\rm bottom}$.
Furthermore, the RS conductance peak feature can be significantly
enhanced.  We note in passing that in the intermediate Rashba
coupling regime $2\alpha^2 \simeq gB$ (not shown),\cite{Tang2012}
the zero point energy of the RS peaks will be changed to the subband
bottom of the spin-up branch.


\section{Concluding Remarks}

In conclusion, we have investigated the interplay of the Rashba SO
coupling and the in-plane magnetic field induced Zeeman effect and
its influence on the spin-resolved subband structure forming the
Rashba-Zeeman effect induced energy gap.  Moreover, we have
demonstrated analytically and numerically the subband structure and
the spin-resolved resonant quantum transport properties of a DFG
system in the presence of a Rashba-Zeeman gap.

Manipulating the DFG system and the Rashba parameter in the strong
Rashba regime, $gB < 2\alpha^2 < 4gB$, or in the ultra-strong Rashba
coupling regime,  $\alpha^2 > 2gB$, allows us to investigate various
bound state features.  These resonant transport features in the DFG
controlled n-type quantum channel include resonant states with
various zero point energy in different Rashba coupling regimes,
hole-like bound states below the subband top of the spin-down
branch, and electron-like quasi-bound states at the threshold of the
spin-up branch.  Our theoretical findings paving the way for the
design of RZ-effect based spintronic device.


 \begin{acknowledgments}
This work was supported by the Ministry of Science and Technology,
Taiwan through Contract No.\ MOST 103-2112-M-239-001-MY3, and the
National Science Council, Taiwan under Contracts No.\
NSC100-2112-M-239-001-MY3, No.\ NSC-100-2112-M-009-013-MY2, and No.\
NSC102-2112-M-009-009-MY2, the Icelandic Research and Instruments
Funds, and the Research Fund of the University of Iceland.
 \end{acknowledgments}


\end{document}